\documentclass[aps,pre,reprint,twocolumn, footinbib,superscriptaddress]{revtex4-2}

\usepackage{microtype}
\usepackage{amssymb,amsmath}
\usepackage{graphicx}
\usepackage{color}
\usepackage{listings}
\usepackage{float}
\usepackage{wrapfig}
\usepackage{amsmath}
\usepackage{gensymb}
\usepackage{soul}
\usepackage{caption}
\usepackage{subcaption}
\usepackage{chemmacros}

\usepackage{upgreek}
\usepackage{enumerate} 
\usepackage[linkcolor = blue, citecolor = blue, urlcolor = blue, colorlinks = true]{hyperref}

\usepackage[version=3]{mhchem}
\usepackage{commath,amssymb}
\usepackage{ushort}
\usepackage{multirow}
\usepackage{cleveref}
\usepackage{epstopdf}

\crefname{equation}{Eq.}{Eqs.}
\crefname{figure}{Fig.}{Figs.}
\crefrangeformat{equation}{Eqs.~#3(#1)#4--#5(#2)#6}

\usepackage[separate-uncertainty]{siunitx}
\newcommand{\mM}{\milli\textsc{m}}

\usepackage{hyphenat}
\usepackage{booktabs}

\usepackage{tabularx}
\usepackage{array}
\usepackage{makecell}
\setcitestyle{super,open={},close={}}

\defcitealias{Verweij2020}{{\color{black}Verweij~{\&}~Moerman~\textit{et al.}}\cite{Verweij2020}}

\newcommand{\affleiden}{\affiliation{\small Huygens-Kamerlingh Onnes Laboratory, Leiden University, P.O. Box 9504, 2300 RA Leiden, The Netherlands}}
\newcommand{\contribeq}{\thanks{These authors contributed equally to this work.}}

\begin{document}

\title{{\Large\bf Brownian motion of flexibly-linked colloidal rings}}

\author{Ruben W. Verweij} 
\contribeq
\affleiden
\author{Julio Melio} 
\contribeq
\affleiden
\author{Indrani Chakraborty}
\affleiden
\author{Daniela J. Kraft}
\email[Corresponding email: ]{kraft@physics.leidenuniv.nl}
\affleiden

\date{\today}

\begin{abstract} Ring, or cyclic, polymers have unique properties compared to linear polymers, due to their topologically closed structure that has no beginning or end. Experimental measurements on molecular ring polymers are challenging due to their polydispersity in molecular weight and the presence of undesired side products such as chains. Here, we study an experimental model system for cyclic polymers, that consists of rings of flexibly-linked micron-sized colloids with $n$=4..8 segments. We characterize the conformations of these flexible colloidal rings and find that they are freely-jointed up to steric restrictions. We measure their diffusive behavior and compare it to hydrodynamic simulations. Interestingly, flexible colloidal rings have a larger translational and rotational diffusion coefficient compared to colloidal chains. In contrast to chains, their internal deformation mode shows slower fluctuations for $n\lesssim 8$ and saturates for higher values of $n$. We show that constraints stemming from the ring structure cause this decrease in flexibility for small $n$ and infer the expected scaling of the flexibility as function of ring size. Our findings could have implications for the behavior of both synthetic and biological ring polymers, as well as for the dynamic modes of floppy colloidal materials.  
\end{abstract}
 
\maketitle

\section{Introduction}

Contrary to their well-known linear cousins, cyclic or ring polymers form a closed structure, without a beginning or an end. Their special topology imparts them with a unique set of properties different from those of linear polymers.\cite{McLeish2005} The absence of free ends suppresses reptation and results in different diffusive properties of ring polymer melts compared to melts of linear polymers, \cite{Halverson2011} and the emergence of a kinetically arrested, glassy state.\cite{Michieletto_2016} Blends of looped and linear polymers display rich viscoelastic properties\cite{kapnistos2008unexpected} and different dynamic and rheological properties from pure melts of only linear or cyclic polymers.\cite{halverson2012rheology} Studying the properties and behavior of ring polymers is not only interesting from a physicists' perspective, but can also be used for the design of materials\cite{liu2020dna} with multifunctional and switchable properties. Understanding the impact of the topological constraint imposed by their ring-nature is also important on the single polymer level, where it might help shed light on how genomes fold themselves into volumes whose linear dimensions are many orders of magnitude smaller than their contour lengths.\cite{Halverson_2014} In particular, their cyclic nature has been proposed to induce stronger excluded volume effects on the conformational and diffusive behavior of the rings than in linear polymers.\cite{prentis1982spatial}

To gain a better understanding of how topology induces these effects, \textit{in situ} measurements of the diffusive dynamics and conformations of single ring polymers are needed. Experimental measurements of their diffusive behavior have focused on ring polymers both as isolated molecules in solution\cite{Serag2014,Robertson2006,ye_interfacial_2016} as well as in melts \cite{richter2015celebrating}. However, these techniques cannot provide the required high spatial and temporal resolution simultaneously that is needed to observe both the dynamics and conformations of individual ring polymers at the single molecule level. Bulk measurements are further complicated by the inherent polydispersity in the molecular weight of the cyclic polymers and the presence of undesired side products stemming from the synthesis.\cite{Haque2020} Despite these challenges, the diffusion coefficient could be obtained experimentally, for example by single-molecule spectroscopy of polymers containing a fluorophore in the presence of linear polymers.\cite{Habuchi2010} Using similar techniques, studies found a larger diffusion coefficient for individual cyclic polymers in solution compared to linear polymers of similar size \cite{Serag2014,ye_interfacial_2016,Robertson2006}. The scaling of the diffusion constant with molecular weight was furthermore found to not be dependent on topology for polymers in solution.\cite{Robertson2006} In contrast, a study on adsorbed polymers at an interface reports a scaling that is topology-dependent, and was attributed to the fact that the diffusion of cyclic polymers is hindered by surface asperities whereas linear polymers are not. \cite{ye_interfacial_2016} In general, there is a difference between the diffusive behavior of ring and linear polymers, but a direct observation of both the conformations and (short-time) diffusive behavior is lacking.
  
To circumvent this limitation, we here use flexible rings made of colloidal particles as model systems instead. Because of their unique combination of microscopic size and their sensitivity to thermal fluctuations,\cite{Einstein1905a,Sutherland1905,Perrin1909} their diffusive properties can be directly studied using well-established techniques, such as optical microscopy, that provide full information of their conformation with high time resolution. Their slower dynamics compared to real molecules make it also possible to investigate loops or chains that consist of fewer repeating units. For example, for reconfigurable colloidal rings built from patchy particles bound via critical Casimir interactions, it was found recently that colloidal analogues of cyclopentane show similar conformational transformations as their atomic counterparts.\cite{swinkels2021revealing} Due to their limited reconfigurability, however, they are unsuitable to serve as analogues for ring polymers. 
In contrast, flexible structures built from spherical colloid-supported lipid bilayers (CSLBs)\cite{VanDerMeulen2013,Rinaldin2019} were found to be completely freely-jointed up to steric exclusions.\cite{Verweij2020,Verweij2021chains} 
We therefore expect a CSLB-based model system to be able to more closely resemble the behavior of ring polymers. 

In this work, we study experimentally and numerically a model system of micron-sized colloidal rings, built from CSLBs, to obtain a detailed understanding of the conformational and diffusive properties of flexible rings. We consider rings of four to eight spherical particles and study both their conformations and diffusive behavior. We find that while the smaller rings show no preferred conformations, preferences arise for the larger rings, because of the increase in degrees of freedom in combination with steric constraints, stemming from the fact that the particles cannot interpenetrate. Both the translational and rotational diffusivity of the rings is greater than that of chains of the same size, because of their smaller radius of gyration. Interestingly, their flexibility, which characterizes the rate of mean squared conformational changes, is lower than that of chains. We show that constraints stemming from the ring structure cause this decrease in flexibility and infer the expected scaling of the flexibility as function of ring size. Our findings may have implications for the behavior of both synthetic and biological ring polymers, as well as for the dynamic modes of floppy colloidal materials.

\section{Materials and Methods}

\subsection{Experimental} 

Flexible colloidal rings were assembled from colloid-supported lipid bilayers (CSLBs).\cite{VanDerMeulen2013,Rinaldin2019} We followed the preparation protocol for the CSLBs as described by \citetalias{Verweij2020}. We employed \SI{2.12\pm0.06}{\um} silica particles as supports, that were coated with a fluid lipid bilayer by deposition and rupture of small unilamellar vesicles consisting of \SI{98.8}{\mole\percent} of the phospholipid DOPC (\iupac{($\Delta$9\hyp{}Cis)
1,2\hyp{}di|o|le|oyl\hyp{}sn\hyp{}gly|ce|ro\hyp{}3\hyp{}phos|pho|cho|line}), with \SI{1}{\mole\percent} of the lipopolymer DOPE\hyp{}PEG(2000) (\iupac{1,2\hyp{}di|o|le|oyl\hyp{}sn\hyp{}gly|ce|ro\hyp{}3\hyp{}phos|pho|e|tha|nol|a|mine\hyp{}N\hyp{}[me|thox|y(po|ly|e|thy|lene
gly|col)\hyp{}2000]}) and \SI{0.2}{\mole\percent} of the fluorescently\hyp{}labeled TopFluor\hyp{}Cholesterol (\iupac{3\hyp{}(di|pyr|ro|me|thene|bo|ron di|flu|o|ride)\hyp{}24\hyp{}nor|cho|les|te|rol}) or, alternatively, the same amount of the fluorescently\hyp{}labeled DOPE\hyp{}Rhodamine
(\iupac{1,2\hyp{}di|o|le|oyl\hyp{}sn\hyp{}gly|ce|ro\hyp{}3\hyp{}phos|pho|e|tha|nol|a|mine\hyp{}N\hyp{}(lis|sa|mine|rho|da|mine
B sulf|o|nyl)}). Bilayer coating was performed in a buffer at pH 7.4 containing \SI{50}{\mM} sodium chloride (NaCl) and \SI{10}{\mM} \iupac{4\hyp{}(2\hyp{}Hy|dro|xy|e|thyl)\hyp{}1\hyp{}pi|per|a|zine|e|thane|sul|fo|nic}
acid (HEPES). We added double-stranded DNA (of respectively strands DS-H-A and
DS-H-B, see the Supplementary Information of \citetalias{Verweij2020}) with an 11 base pair long sticky end and a double stearyl anchor, which inserts itself into the bilayer via hydrophobic interactions, as shown \cref{LOfig:fig1}a. The sticky end of strand DS-H-A is complementary to the sticky end of strand DS-H-B, which allows them to act as linkers. DNA hybridization and experiments were performed in a different buffer of pH 7.4, containing \SI{200}{\mM} NaCl and \SI{10}{\mM} HEPES. Rings of \SI{2.12}{\um} CSLBs were formed by self-assembly or via manual assembly using optical tweezers, in a sample holder made of \iupac{po|ly|a|cryl|a|mide} (PAA) coated cover glass.\cite{Verweij2020}
Confocal microscopy images of the coated particles are shown in
\cref{LOfig:fig1}c for a tetramer loop, \cref{LOfig:fig1}d for a hexamer loop and \cref{LOfig:fig1}e for an octamer loop.

\subsection{Microscopy}

Loops were imaged for at least \SI{5}{\minute} (frame rates
between 5 and \SI{20}{fps}) at room temperature using an inverted confocal
microscope (Nikon Eclipse Ti-E) equipped with a Nikon A1R confocal scanhead
with galvano and resonant scanning mirrors. A $60\times$ water immersion
objective (NA=1.2) was used. 488 and \SI{561}{\nm} lasers were used to excite,
respectively, the TopFluor and Rhodamine dyes. Laser emission
passed through a quarter wave plate to avoid polarization of the dyes and the
emitted light was separated by using $500-\SI{550}{\nm}$ and
$565-\SI{625}{\nm}$ filters.

To complement the data obtained from self-assembled loops, we used optical
tweezers to assemble specific cluster sizes. For the hexamer and octamer loops,
the probability of forming such a loop using the self-assembly method we used
here is low, therefore these were formed exclusively using optical tweezers.
Briefly, we employed a homemade optical setup consisting of a highly focused
trapping laser manufactured by Laser QUANTUM (\SI{1064}{\nm} wavelength). The
laser beam entered the confocal microscope through the fluorescent port, after
first passing through a beam expander and a near-infrared shortpass filter. The
same objective was used for imaging and to focus the trapping laser beam.
During the trapping, the quarter wave plate was removed from the light path.

Particle positions were tracked using a custom algorithm\cite{Rinaldin2019}
available in TrackPy by using the \texttt{locate\_brightfield\_ring} function
\cite{trackpy} or using a least-square fit of a Mie scattering based model
implemented in HoloPy.\cite{HoloPy} Both methods agree to an accuracy of at
least \SI{1}{px}, however we have found that the Mie scattering based model is
more robust for tracking multiple particles in close proximity to each other.
For all analysis, we only selected rings that showed all bond
angles during the measurement time, experienced no drift and were not stuck to
the substrate.  An overview of the total number of measurements, the total
duration and the total number of frames per ring size is shown in
\cref{LOtable:data}.

\begin{table}
    \caption{Overview of the number of measurements, the total duration and the total number of frames per ring size, for the experimental and simulated data.\label{LOtable:data}}
\smallskip\noindent
\centering\sisetup{table-number-alignment=center}
    \begin{tabularx}{\linewidth}{X S[table-format=3] S[table-format=2] S[table-format=3] S[table-format=4] S[table-format=1.1e1] S[table-format=1.1e1]}
        \toprule
        {\thead[l]{$n$}} & \multicolumn{2}{c}{\thead{Measurements}} & \multicolumn{2}{c}{\thead{Duration [min]}} & \multicolumn{2}{c}{\thead{Total frames}} \\
                    & {\footnotesize Exp.} & {\footnotesize Sim.} & {\footnotesize Exp.} & {\footnotesize Sim.} & {\footnotesize Exp.} & {\footnotesize Sim.} \\ \midrule
        {4} & 14  & 20 & 196 & 600 & 2.2e5 & 2.5e7 \\
        {5} &    & 20 &     & 600 &       & 2.5e7 \\
        {6} & 10  & 20 & 141  & 600 & 1.7e5 & 2.5e7 \\
        {7} &    & 20 &     & 600 &       & 2.5e7 \\
        {8} & 9  & 20 & 110  & 600 & 1.3e5 & 2.5e7 \\
        \bottomrule
    \end{tabularx}
\end{table}

\subsection{Simulations}\label{LOsec:simulations}

We have performed Brownian dynamics simulations with hydrodynamic interactions
following the method outlined in \citet{Sprinkle2020} using the open-source
RigidMultiblobsWall package.\cite{RigidMultiblobsWall} The procedure is
identical to the method described in \citet{Verweij2021chains}, which we now briefly
summarize. Hydrodynamic interactions are calculated using the Stokes equations
with no-slip boundary conditions. The hydrodynamic mobility matrix is
approximated using the Rotne-Prager-Blake (RPB) tensor,\cite{Swan2007} which
is a modified form of the Rotne-Prager-Yamakawa (RPY)
tensor\cite{Rotne1969,Yamakawa1970,Wajnryb2013} and accounts for a bottom
wall, which is unbounded in the transverse directions. These corrections to the
RPY tensor are combined with the overlap corrections described in
\citet{Wajnryb2013} to prevent particle-particle and particle-wall overlap.
The RPB mobility inaccurately describes near-field hydrodynamic interactions
and therefore breaks down for small separation distances. This can be overcome
by adding a local pairwise lubrication correction to the RPB resistance matrix as described in detail in \citet{Sprinkle2020} Based on the full lubrication-corrected hydrodynamic mobility matrix, the Ito overdamped Langevin equation is solved to describe the effect of thermal fluctuations.

We include a gravitational force on the particles to confine them to diffuse close to the bottom wall, as in the experiments. Inter-particle bonds are modeled by harmonic springs of stiffness $1000 k_{B}T/R^2$ and equilibrium
length $2R$, where $R=\SI{1.06}{\um}$ is the particle radius. The bond angle is
not restricted. We set the temperature $T=\SI{298}{\kelvin}$, the viscosity of
the fluid $\eta=\SI{8.9e-4}{\pascal\second}$, the gravitational acceleration
$g=\SI{9.81}{\meter\per\second\squared}$, the particle mass $m_{p} =
\SI{9.5e-15}{\kilo\gram}$ (by assuming a particle density of
\SI{1900}{\kilo\gram\per\cubic\meter}) and the simulation timestep $\Delta t =
\SI{1.42}{\milli\second}$. For the firm potential that prevents overlap, we use
a strength of $4 k_{B}T$ and a cutoff distance\cite{Usabiaga2017,Sprinkle2020}
$\delta_{\mathrm{cut}} = \num{e-2}\,R$. We initialized the particle loops in
the configuration given by the regular polygon of the same size. Then, these
initial configurations were randomized by running the integration for a
simulated time of \SI{60}{\second} prior to saving the configurations, to
ensure a proper equilibration of the particle positions, bond lengths,
velocities and opening angles. The particle positions were saved every 8
simulation steps to obtain a final framerate of approximately \SI{90}{fps}.  An
overview of the total number of simulations, the total duration and the total
number of saved frames per loop size is shown in \cref{LOtable:data}.

For comparison to the simulated and experimental data, we have generated
permutation data, in which the rings are completely non-interacting and
freely-jointed up to steric exclusions. Firstly, we have generated all
permutations of the $(n-3)$ independent opening angles $\theta_i$, each of
which can take on $N_{\theta}=(360-2\times60)/(\delta\theta)$ different values,
where $\delta\theta$ denotes the bin width. The opening angles are indicated by
the blue arcs in the schematics of \cref{LOfig:fig1B} for all ring sizes. This
gives a total number of $P(N_{\theta},n-3) = N_{\theta}! / (N_{\theta} -
(n-3))!$ combinations of the $(n-3)$ opening angles $\theta_i$. Secondly, we
removed those combinations that are forbidden because of steric exclusions
between particles.  After removing these configurations, we checked if the
topology of the structure was correct and removed configurations of the wrong
topology, i.e.\ structures that did not form a closed ring. This resulted in
the final allowed combinations, which we call ``permutation data''.

\subsection{Diffusion tensor analysis}\label{LOsec:diffusiontensordef}

\subsubsection{Definition of the diffusion tensor}

We have determined the short-time diffusivity of the rings, both as function of
their instantaneous shape for the tetramer rings, as well as averaged over all
possible configurations for all loop sizes. Because the rings are sedimented to
the bottom substrate, we consider only the quasi-2D, in-plane diffusivity.

For all rings, we have determined the short time diffusion tensor,
\begin{align} \bm{D}[ij] &\equiv \frac{1}{2n} \frac{\partial\langle\Delta i
\Delta j \rangle_\tau}{\partial \tau}, \label{LOeq:dtensor} \end{align} with
$\tau$ the lag time between frames, $\langle \cdots \rangle_\tau$ denotes a
time average over all pairs of frames $\tau$ and $\Delta i = i(t+\tau) -
i(t)$ is the displacement of the $i$-th diffusion tensor element. The average
diffusion tensor elements $\bm{D}[ij]$ were obtained by fitting the overall
slope of the mean (squared) displacements as a function of lag time $\tau$. We
considered lag times up to \SI{0.25}{\second}, set by the frame rate of the
experimental data. For fitting the slopes of the MSDs, we used a standard least-square 
fit of a linear model with an intercept, in order to characterize the localization error.\cite{michalet2010mean}

Using \cref{LOeq:dtensor}, we have calculated the shape-averaged, quasi-2D
translational diffusion coefficient corresponding to in-plane diffusivity above
the wall and given by $D_T = \bm{D}[\bm{r}_{t.p.}\bm{r}_{t.p.}]$, where
$\bm{r}_{t.p.}$ is the position of the loop relative to the tracking point
defined by \cref{eq:tp}. Additionally, we determined the rotational diffusion
coefficient $D[\alpha\alpha]$ from the mean squared angular displacement of the
$x$-axis (defined in \cref{LOeq:xaxis}, as depicted schematically in \cref{LOfig:fig2}a for
the tetramer loop), which describes the rotational diffusivity
around an axis perpendicular to the substrate. Finally, we determine the
overall cluster flexibility $D[\bm{\theta\theta}]$ by calculating the mean
squared displacements of all $n$ opening angles $\theta_i$ as follows:
\begin{align} \langle \lvert\bm{\Delta\theta}\rvert^2 \rangle &=
\langle \lvert \left( \Delta\theta_1, \dots, \Delta\theta_{n}
\right) \rvert^2 \rangle, \label{LOeq:deltatheta}\end{align} so
that the flexibility $D[\bm{\theta\theta}]$ is given by
\begin{align} \langle \lvert\bm{\Delta\theta}\rvert^2 \rangle &= 2
n D[\bm{\theta\theta}] t,\label{LOeq:flexibility} \end{align}
analogously to the other diffusion tensor elements.

\subsubsection{The definition of the coordinate system}\label{LOsec:trackingpoint}

As tracking point, we have used the center of diffusion (c.d.) which coincides
with the center of mass of colloidal rings. This is an important choice because
it affects the magnitude of the diffusion
tensor.\cite{Wegener1985,Cichocki2019} The coordinate system used here is
identical to the coordinate system described in \citet{Verweij2021chains} and
we briefly summarize its definition here. The c.d. was calculated from
$\bm{A}_{ij}$ defined by Equation 2.16 of \citet{Cichocki2019} using the RPB
tensor\cite{Swan2007} with lubrication corrections as the inter-particle
mobility matrix $\bm{\mu}_{ij}$.  This tensor includes wall corrections, as
discussed previously in~\cref{LOsec:simulations}. The c.d. was determined from
the simulated particle positions, because the height above the bottom wall was
not measured experimentally, but is needed to calculate the wall corrections.
The direction of the body-centered $x$- and $y$-axes was determined as function
of the tracking point $\bm{r}_{t.p.}$, which defines the origin of the
body-centered coordinate frame. We define 
\begin{align}\bm{r}_{t.p.} &= \rho_1 \bm{r}_1 + \rho_2 \bm{r}_2 + \dots + \rho_n \bm{r}_n\label{eq:tp}\end{align}, which defines the location of the
tracking point as a linear combination of the particle positions (Equation 2.2
and 2.3 of \citet{Cichocki2019}). $\bm{\rho} = (\rho_1, \rho_2, \dots, \rho_n)$
is a weight vector which determines how much weight is accorded to each
particle in the calculation of the tracking point $\bm{r}_{t.p.}$. As an
example, for a tetramer ring, $\bm{\rho} = (1/n = 1/4, 1/4, 1/4, 1/4)$ when the
tracking point is chosen to be the c.d., or equivalently, the center of mass.

The direction of the $x$-axis was chosen as \begin{align} \bm{\hat{x}} &=
\pm\left[\dfrac{\bm{r}_{t.p.,1} + \dots + \bm{r}_{t.p.,s_1}}{\rho_{1} +
        \dots + \rho_{s_1}} - \dfrac{\bm{r}_{t.p.,s_2} + \dots +
        \bm{r}_{t.p.,n}}{\rho_{s_2} + \dots +
\rho_{n}}\right],\label{LOeq:xaxis} \end{align} where $\bm{r}_{t.p.,i}$ is
the $i$-th coordinate of the tracking point and the loop is split
into two parts with equal numbers of particles according to \begin{align}
    \begin{cases} s_1 = s_2 = \lceil \frac{n}{2} \rceil & \text{for odd } n
    \\ s_1 = \lceil \frac{n}{2} \rceil, s_2 = s_1 + 1 & \text{for even } n
\end{cases} \end{align} $\bm{\hat{y}}$ is then chosen such that
$\bm{\hat{x}}$ and $\bm{\hat{y}}$ form a right-handed coordinate system,
where the direction of $\bm{\hat{y}}$ is chosen to point along
$\bm{r}_{t.p.} - \left(\bm{r}_{s_1} + \bm{r}_{s_2}\right)/2$.  This
orientation was determined for every frame, which fixed the orientation of
the body-centered coordinate system $\bm{x}(\tau=0), \bm{y}(\tau=0)$. For
subsequent lag times, the direction of $\bm{y}(\tau)$ was chosen such that
$\bm{y}(\tau=0)\cdot \bm{y}(\tau) > 0$, i.e.\ the direction of $\bm{y}$ does
not change sign. The resulting coordinate system relative to the c.d. is
visualized for the tetramer loops in \cref{LOfig:fig2}a.

\subsubsection{Shape-dependent diffusivity of the tetramer loops}

For the flexible tetramer loops, in addition to the \emph{shape-averaged} short
time diffusion tensor, we have calculated the \emph{shape-dependent} short time
diffusion tensor. This is feasible for the tetramer loop because its shape is
fully characterized by using only one opening angle. To do so, we have
calculated a $4\times4$ diffusion tensor, where the four degrees of freedom
correspond to translational diffusivity in $x$ and $y$, rotational diffusivity
and the flexibility of the tetramer loop, which is described by the diffusivity
of the opening angles $\theta_{i}$, which we have defined in
\cref{LOeq:flexibility}.  Specifically, for the tetramer loop, the $x$- and
$y$-directions are schematically shown for one configuration in
\cref{LOfig:fig2}a and defined by \cref{LOeq:xaxis}. The rotation angle used
for determining the rotational diffusivity is indicated in \cref{LOfig:fig2}a
and is the angle of the $x(\tau)$ relative to $x(\tau=0)$, i.e.\ the angle of
the body-centered $x$-axis of the current frame relative to the body-centered
$x$-axis of the reference frame at $\tau=0$. The flexibility is calculated from
the mean-squared displacement of the opening angle $\theta$.  As illustrated by
the blue arcs in the schematic of \cref{LOfig:fig1B}b, $\theta$ is defined in
such a way that it is always less than or equal to \SI{120}{deg}.

The diffusion tensor elements of the tetramer loops were determined analogously
to the trimers.\cite{Verweij2020} Briefly, for each pair of frames, we
determined the initial shape of the ring, which is characterized by the opening
angle $\theta$. We only considered trajectories where the variation in $\theta$
did not exceed the edges of the bin describing the initial shape. That is, we
divided the possible values of $\theta$ in bins and calculated the short-time
diffusivity given by \cref{LOeq:dtensor} for all combinations of lag times
where $\theta(\tau)$ remained in the same bin as $\theta(0)$, which were then
stored according to their respective $\theta$-bins. In that way, we calculated
the diffusion tensor elements separately for each initial shape. 

\begin{figure} \centering     \includegraphics{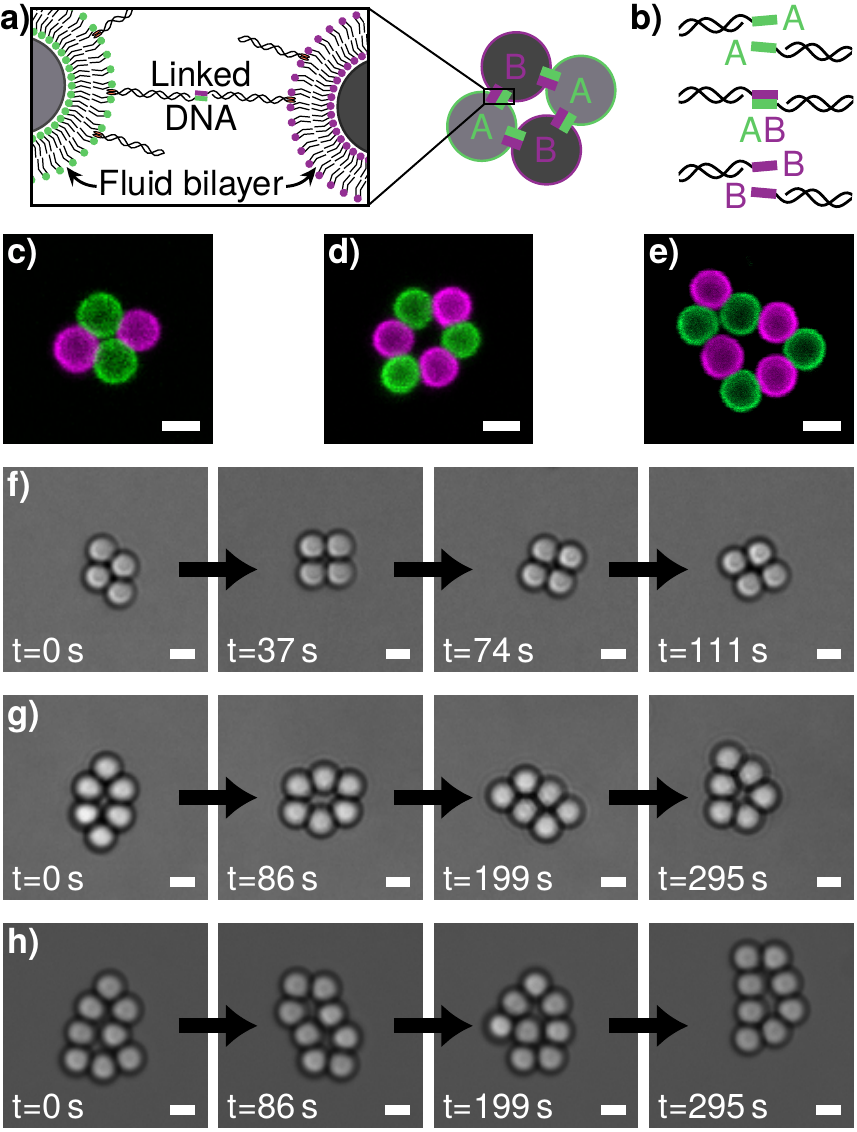}
    \caption{\textbf{Flexibly-linked colloidal loops.} \textbf{a)} The flexible
        rings are built from colloid-supported lipid bilayers (CSLBs). CSLBs
        consist of spherical silica colloids coated with a fluid lipid bilayer.
        DNA linkers are inserted into the bilayer using a hydrophobic anchor.
        Because of the fluid lipid bilayer, the linkers can diffuse on the
        surface of the particles and therefore, the particles can move with
        respect to each other whilst staying bonded. \textbf{b)} The DNA
        linkers are functionalized with sticky ends A that are complementary to
        sticky ends B, so that particles functionalized with A-type linkers can
        only form bonds with particles coated with B-type linkers.  \textbf{c-e)}
        Confocal images of a tetramer (c, $n=4$), hexamer (d, $n=6$) and octamer (e, $n=8$) loop. \textbf{f-h)} Bright field snapshots of a flexible tetramer (f, $n=4$), hexamer (g, $n=6$) and octamer (h, $n=8$) ring, which show shape changes that are more pronounced for the larger rings. Scalebars in panels c-h are \SI{2}{\um}.\label{LOfig:fig1}} \end{figure}

\section{Results and Discussion} 

\subsection{Free energy of different conformations of flexible loops}\label{LOsec:energy_loops}

We designed our colloidal model system for ring polymers by assembling flexible rings of colloid supported lipid bilayers (CSLBs), i.e.\ colloidal silica particles surrounded by a fluid lipid bilayer.\cite{VanDerMeulen2013,Rinaldin2019} We equip the CSLBs with strong and specific bonds imparted by DNA linkers with single stranded sticky ends that are inserted into the lipid bilayer. The DNA linkers are mobile on the CSLB surface allowing for configurational flexibility of structures assembled from bonded CSLBs also after assembly.\cite{Chakraborty2017, Verweij2020, Verweij2021chains} 
 We employed two types of DNA linkers with complementary sticky ends, which we label A and B, such that A only binds to B and not to itself, see \cref{LOfig:fig1}b. 
 The colloidal size of our model system allows us to observe the position of the constituent spheres using bright field microscopy. We visualize their functionalization with the two strands as well as their correct and selective binding by integrating dyes with the DNA linkers, and imaging the rings with confocal microscopy, see \cref{LOfig:fig1}c, d and e. The use of two complementary sticky ends A and B reduces, or, in the case of a tetramer loop, even prevents adhesion between opposing particles in the ring and hence a change of its topology and reconfigurability. This strategy limits our experiments to even-numbered ring sizes, of which we assembled tetramers ($n=4$), hexamers ($n=6$) and octamers ($n=8$).  We therefore compared and complemented our experiments with Brownian dynamics simulations for colloidal rings with $n=4-8$ constituent spheres. In these simulations, hydrodynamic interactions between particles and the substrate were taken into account via the Rotne-Prager-Blake (RPB) tensor,\cite{Swan2007} overlap corrections\cite{Wajnryb2013} and a local pairwise lubrication correction\cite{Sprinkle2020} (see \cref{LOsec:simulations} for details). 

\begin{figure*} \centering \includegraphics{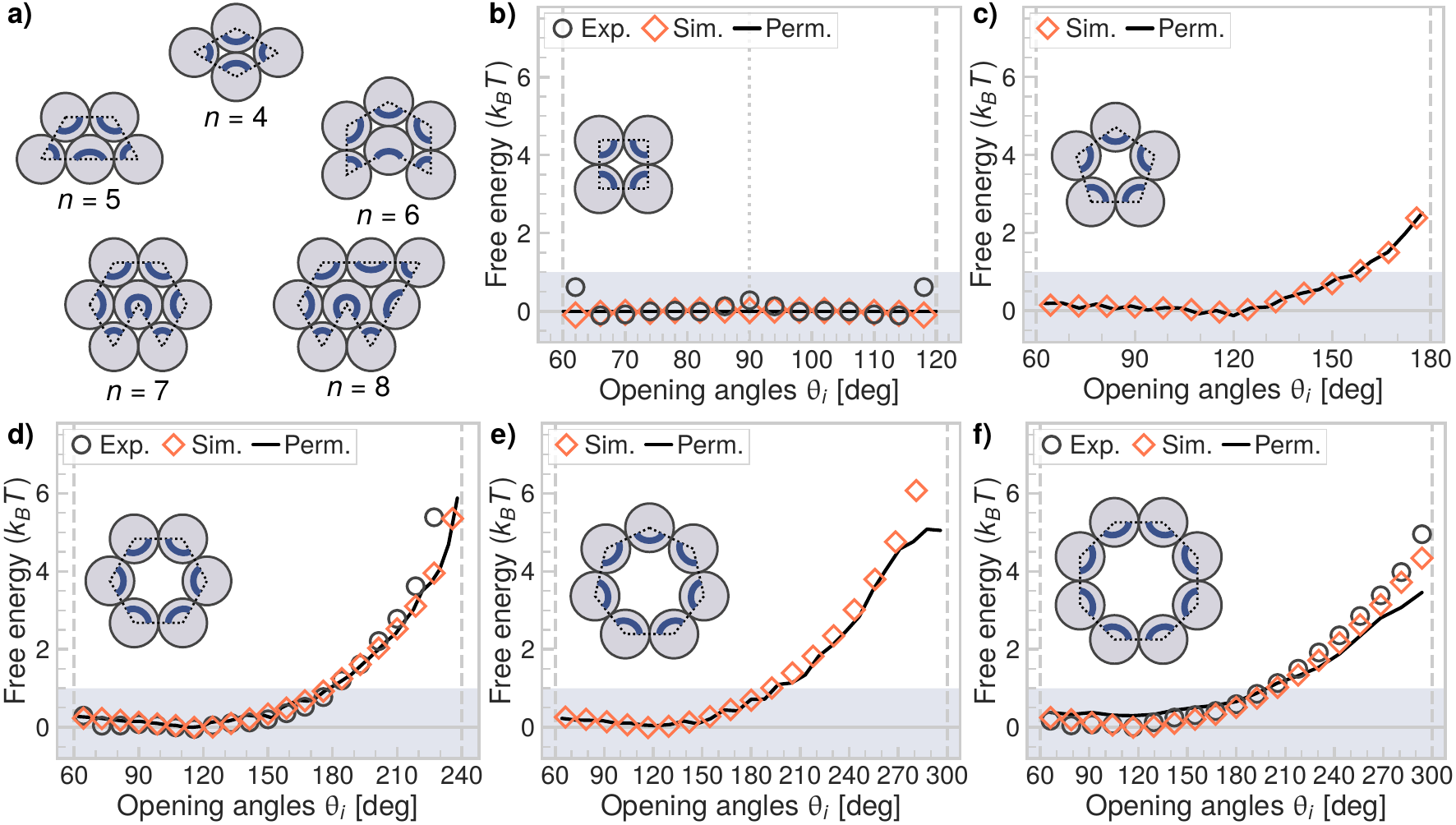}
    \caption{\textbf{Conformations of flexibly-linked colloidal loops.}
        \textbf{a)} The most compact configurations for all rings studied here, from $n=4-8$. The opening angles $\theta_{i}$ are indicated by the blue arcs. \textbf{b-f)} The free energy of \textbf{b)} tetramer ($n=4$), \textbf{c)} pentamer ($n=5$), \textbf{d)} hexamer ($n=6$), \textbf{e)} heptamer ($n=7$) and \textbf{f)} octamer ($n=8$) rings, for $\circ$ experimental, $\diamond$ simulated and permutation data (black line). The shaded area indicates free energy differences lower than the thermal energy. Blue lines indicate internal opening angles taken into account. The free energy was determined using Boltzmann weighing of the joint probability density function of all opening angles $\theta_{i}$.
        \label{LOfig:fig1B}}
\end{figure*}

The thus assembled colloidal rings show constant shape changes induced by thermal fluctuations. The full flexibility of the rings is clearly visible in the representative time series from bright field microscopy movies of flexible tetramers, hexamers, and octamers, see \cref{LOfig:fig1}f-h and corresponding movies in the Supporting information. Tetramers possess only one internal degree of freedom and randomly transition between diamond-like configurations with internal opening angles $\theta_{i}$ between $\SI{60}{deg}\leq \theta_i \leq \SI{120}{deg}$. With increasing ring size, the number of degrees of freedom rises and the constrained motion observed in the tetramer is gradually lifted.  Hexamers and octamers possess three and five internal degrees of freedom, respectively, and can adopt an increasingly wider range of shapes. Correspondingly, the maximum internal opening angle that is geometrically possible increases from $\SI{180}{deg}$ for $n=5$, to  $\SI{240}{deg}$ for $n=6$ and to $\SI{300}{deg}$ for $n\geq 7$, the maximum that is achievable for three otherwise unconstrained particles in a chain (see \cref{LOfig:fig1B}a). At $n=7$ a compact hexagonal structure is required for attaining this maximum internal angle. Additional particles beyond $n=7$ in the ring do not increase the maximum internal opening angle, but the number of conformations of the remaining particles in case that any three take on the maximum internal angle is higher. In turn, this increases the probability to observe a configuration with the maximum value of the opening angle. 

To quantify the conformations of the colloidal rings, we measured all $n$ indistinguishable internal opening angles $\theta$ of a ring of size $n$ and calculated the probability density function $p (\theta)$. From this, we determined the free energy $F$ using Boltzmann weighing, \begin{align} \frac{F}{k_BT} &= -\ln{p(\theta)} +\frac{F_0}{k_BT} \label{LOeq:boltzmann}, \end{align} where $k_B$ is the Boltzmann constant, $T$ the temperature,  and $F_0$ is a constant and arbitrary offset to the free energy. The angular range is determined purely by geometry: the minimum opening angle is $\theta=\SI{60}{deg}$ and the maximum angle depends on $n$ and is limited by the constraints induced by the topology and geometry, as shown in \cref{LOfig:fig1B}a and ranges from $\theta=\SI{120}{deg}$ for $n=4$ to $\theta=\SI{360}{deg}$ for $n\geq 7$ as noted before. 

We plot the obtained values for the free energy as a function of the possible range of opening angles for rings of $n=4-8$ in \cref{LOfig:fig1B}. For all datasets, we set the free energy to 0 at the smallest value and compare it to the thermal energy of $1~k_{B}T$, which is indicated by the shaded area. The tetramer rings show no preference for any opening angle with respect to the thermal energy. They transition freely and with equal probability between all possible internal opening angles, as shown in \cref{LOfig:fig1B}b. For the pentamer loops, however, the free energy exceeds the thermal energy for large internal opening angles $\theta_{i}\geq\SI{158}{deg}$, see \cref{LOfig:fig1B}c. Similarly, for $n=6-8$ we also find that the free energy exceeds the thermal energy for increasingly greater values of the opening angles: for hexamer loops this occurs if $\theta_{i}\geq\SI{178}{deg}$, for heptamer loops if $\theta_{i}\geq\SI{192}{deg}$, and for octamer loops if $\theta_{i}\geq\SI{204}{deg}$. This finding implies that smaller opening angles occur more frequently, leading to an effective free energy preference, and is found both in experimental measurements and simulations.  

We can attribute this free energy preference to the steric constraints imposed by the ring topology and self-avoidance of the segments of the ring. To demonstrate this, we count the number of possible conformations of the ring for a given opening angle taking steric constraints into account. We then extract the free energy from this permutation data and plot it alongside the experiments and simulations in \cref{LOfig:fig1B}. The good agreement clearly confirms that these preferences purely arise from steric constraints: there are simply fewer possible configurations that include large values of the opening angles. Indeed, for $n=4-7$ compact structures of the ring are required to observe the largest opening angles, as depicted in \cref{LOfig:fig1B}a, with concomitantly fewer possible conformations of the rings. In fact these steric constraints defined the maximum of the opening angles in the first place. Comparing the difference between the minimum and maximum values of the free energy between different ring sizes we see that the difference increases with $n$ until $n=7$. This implies that the ratio between the observable configurations for one angle in the maximum angle range and those for any other value of the opening angle decreases. For $n=8$, the steric constraint is partially lifted and more conformations are possible for one internal angle at the maximum possible value. As a consequence, the maximum free energy difference lowers again compared to $n=7$.

\subsection{Gyration radii of rings and chains}\label{LOsec:chainsandrings}

\begin{figure*} \centering \includegraphics{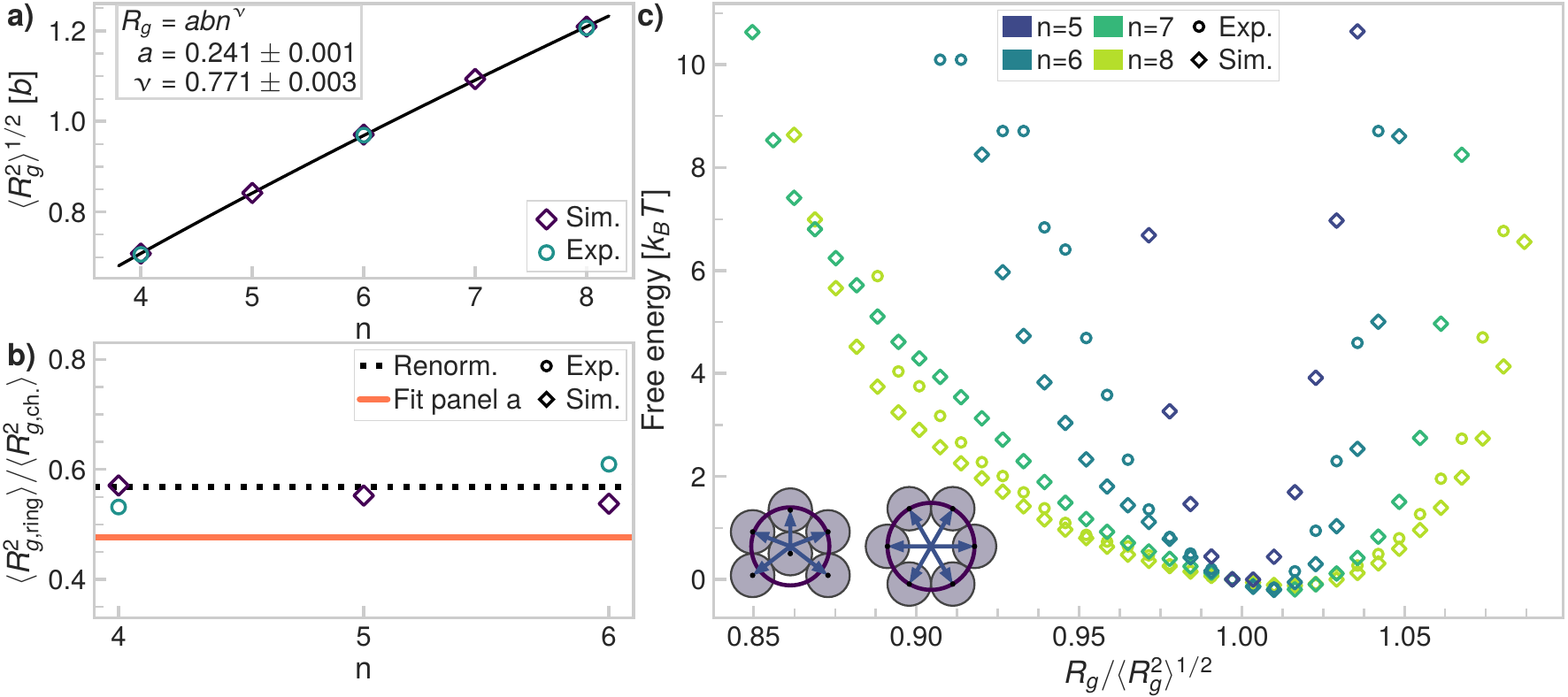}
    \caption{\textbf{Radius of gyration of colloidal rings.} \textbf{a)} The
    radius of gyration in units of the bond length $b$ scales as function of $n$ as predicted by \cref{eq:Rgnscaling}, which is also shown in the inset, along with the values of the fitted parameters. \textbf{b)} The ratio between the radius of gyration of rings and chains is close to the value predicted by Renormalization theory,\cite{prentis1982spatial,hegde2011conformation} as show for the experimental and simulated data that are available for both the rings and chains. Additionally, the solid line indicates the ratio of the scaling found in panel a for rings and the previously determined scaling of chains, which takes into account all available ring ($n=4$ to 8) and chain ($n=3$ to 6) lengths. \textbf{c)} The free energy in terms of the radius of gyration divided by its average value for all ring sizes. The distribution is asymmetric and covers a wider range of possible values for the larger ring sizes. Note that the radius of gyration of tetramer rings is constant and therefore not included here. The schematics show a circle with its radius equal to the radius of gyration of the most compact (left) and most extended (right) hexamer ring.\label{LOfig:fig4}} \end{figure*}

Whilst the opening angles between particles in a chain can be used to uniquely describe its conformation, they are not a very intuitive measure for how compact or extended the structure is. A better means that is also used to quantify the extent of polymer chains and colloidal bead-chains is the radius of gyration $R_g$, which is defined as 
\begin{align} R_{g} = \left[\frac{1}{n}
        \sum^{n}_{i=1} {\left|\bm{r}_{i} - \bm{r}_{c.m.}
\right|^2}\right]^{1/2}, \label{LOeq:Rg} \end{align} where $\bm{r}_i$ is the position of the $i$-th sphere and $\bm{r}_{c.m.}$ is the position of the center of mass of the loop. 

The radius of gyration of rings is expected to follow the same scaling as the radius of gyration of chains\cite{jagodzinski1992universal,prentis1982spatial}, given by:\cite{DeGennes1979} \begin{align} R_{g} &= a b n^{\nu},\label{eq:Rgnscaling} \end{align} with $a$ a positive constant, $b$ is the Kuhn length (approximately equal to the bond length\cite{Verweij2021chains}) and the Flory exponent ${\nu=3/(d+2)} = 3/4$ for a self-avoiding walk in $d=2$ dimensions\cite{kamien_flory_1993, Wang1995}. As shown in \cref{LOfig:fig4}a, we find an excellent agreement between the predicted scaling law and the experimental and simulated data. The fitted value of $\nu=\num{0.771\pm0.003}$ is in close agreement with the theoretically expected value from renormalization theory\cite{kamien_flory_1993, Wang1995} of $3/4$.  This scaling coefficient also agrees with the  value we have previously determined for chains $\nu=\num{0.726\pm0.005}$\cite{Verweij2021chains}, as would be expected since the scaling coefficient should be independent of topology. 

For the scaling constant $a$ however, we find a lower value for the rings ($a_{ring}=\num{0.241\pm0.001}$) compared to the chains\cite{Verweij2021chains} ($a_{chain}=\num{0.349\pm0.002}$). This means that, as intuitively expected, the rings are on average more compact than chains of the same number of particles. We explore the ratio between $R_g^2$ of the rings and chains $G=R^2_{g,ring}/R^2_{g,chain}$ in greater detail in \cref{LOfig:fig4}b, by comparing the simulated and experimental values to the predicted ratio of $G=\num{0.568}$ based on Renormalization theory,\cite{prentis1982spatial,hegde2011conformation} which is close to the values that we obtain here. Because the chain data is only available for $n=3$ to 6 while for the loops, we consider sizes of $n=4$ to 8, we have also determined $G$ by dividing $a^2_{ring}$ by $a^2_{chain}$. As shown in \cref{LOfig:fig4}b, the value that we
obtain from the fitted scaling relations ($G=0.48$) is also close to the theoretically expected value. Taken together, these results show that our colloidal model system exhibits the same scaling of $R_{g}$ as molecular polymers based on predictions from polymer theory, which is surprising considering their low number of segments $n$.

Finally, in \cref{LOfig:fig4}c, we show the free energy of the rings in terms of their radius of gyration, which we normalized by the average value. Firstly, we note that the agreement between the simulated and experimental data is good for all ring sizes, although we systematically observe higher free energy values at a give value of the normalized radius of gyration than for the experiments. Secondly, it becomes apparent that the distribution of the free energy in terms of the scaled radius of gyration is asymmetric for all ring sizes and that is covers a greater range for the larger ring sizes. This last observation can be intuitively understood to be caused by the increasing number of particles and also, the increasing number of degrees
of freedom. Furthermore, we observe that either very compact or very extended structures are less likely than intermediate structures, with differences in the free energy of the most likely extents compared to the least likely ones reaching up to around $10~k_{B} T$. These differences in the free energy are likely caused by excluded volume interactions in the case of the compact structures and steric constraints to preserve the ring topology for the more extended configurations. 

To summarize, the size of the colloidal rings can be characterized by their radius of gyration, which follows the scaling relations that are predicted from polymer theory. The ring topology and excluded volume interactions play an important role, which may also affect the diffusive properties of the colloidal rings, which we will now consider.

\subsection{Diffusion of flexible tetramer loops} 

\begin{figure*} \centering     \includegraphics{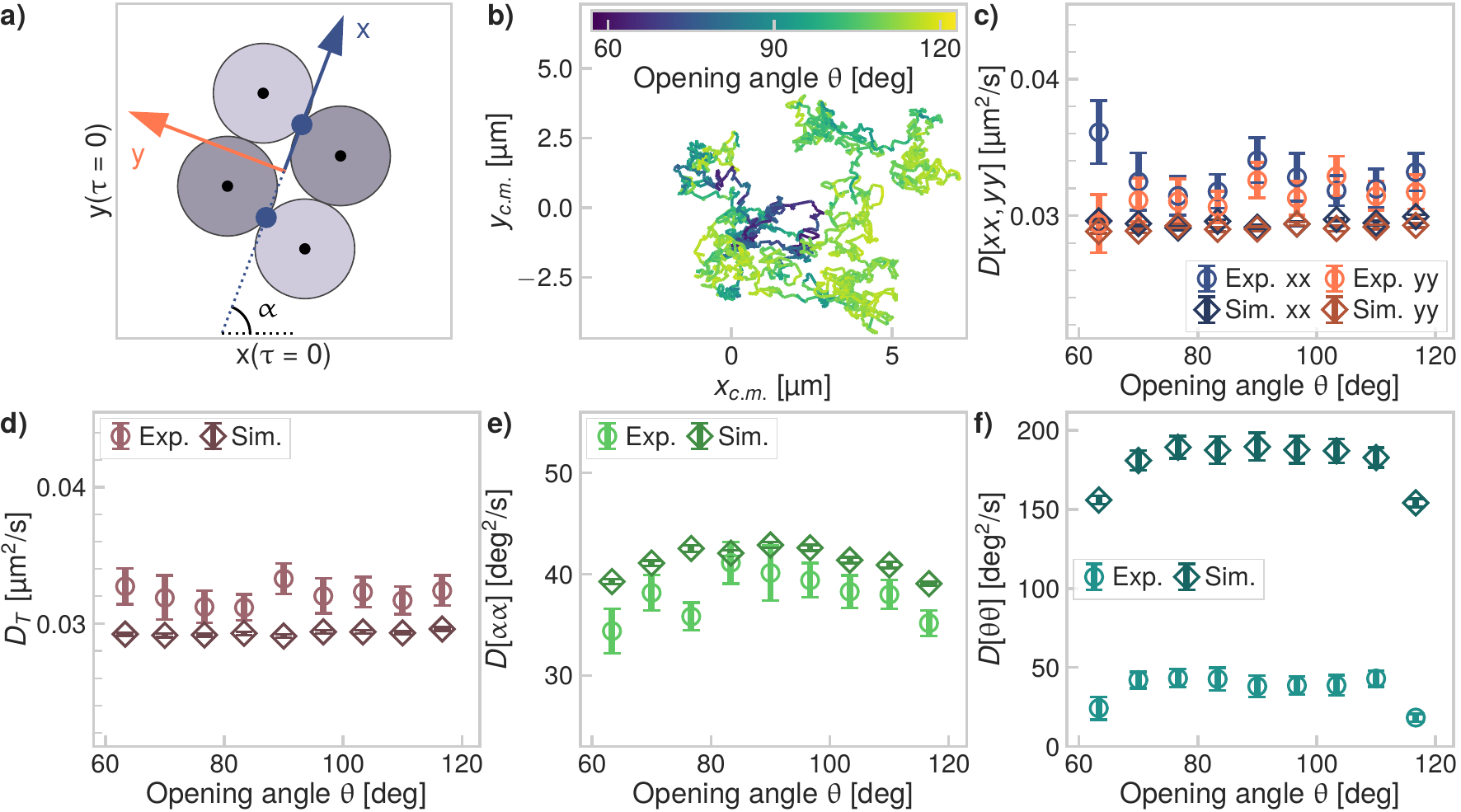}
    \caption{\textbf{Diffusion of flexible tetramer loops.} \textbf{a)} An
        illustration of the coordinate system used to analyze the diffusivity
        of the tetramer rings, as defined in
        \cref{LOsec:diffusiontensordef}. \textbf{b)} The displacement of the center of mass (c.m.) of a \SI{10}{\min} experimental measurement of a tetramer loop. The color indicates the instantaneous value of one of the opening angles $\theta$. \textbf{c-f)} We have compared the
        diffusion tensor elements calculated from $\diamond$ simulated and
        $\circ$ experimental data. \textbf{c)} The translational diffusion along the x- and 
        y-directions is comparable in magnitude and shows little shape dependence.
        \textbf{d)} The in-plane translational
        diffusion coefficient $D_T$. We find that $\left\langle
        D_{\mathrm{exp.}}/D_{\mathrm{sim.}} \right\rangle = \num{1.10\pm0.02}$.
         \textbf{e)} The rotational diffusivity, for which
         $\left\langle D_{\mathrm{exp.}}/D_{\mathrm{sim.}} \right\rangle = \num{0.92\pm0.04}$.  \textbf{f)} Compared to the simulated flexibility, the
        experimental flexibility is much lower, namely $\left\langle
        D_{\mathrm{exp.}}/D_{\mathrm{sim.}} \right\rangle = \num{0.20\pm0.04}$.
        All off-diagonal diffusion tensor elements are close to
zero.\label{LOfig:fig2}} \end{figure*}

The continuous change of the ring's configuration might also affect its short-time diffusive behavior. We will here analyze the diffusive behavior of the simplest ring shape, the tetramer, because of its simplicity and equal probability of all configurations. In \cref{LOfig:fig2}a, we schematically show the coordinate system relative to the center of mass (c.m.) of the tetramer loop, which is defined in \cref{LOsec:diffusiontensordef}. The position of the c.m.\ as function of time for a \SI{10}{\min} experimental measurement is depicted in \cref{LOfig:fig2}b. The variations in color, which
indicate the instantaneous value of one of the opening angles $\theta$, clearly show that indeed, the tetramer loops continuously change their shape while they diffuse.

These continuous shape changes could affect the translational diffusivity of the tetramer loops. However, as shown in \cref{LOfig:fig2}c for the shape-dependent, short-time translational diffusivity with respect to the $x$-
and $y$-axes, we find that the translational diffusivity is constant and does not depend on the opening angle $\theta$. This may be explained by the fact that for tetramer loops, the radius of gyration, which is a measure for their size, is constant as well. Therefore, the translational diffusivity can be well
described in terms of the average diffusivity $D_{T}$ of the $x$- and $y$-axes that is shown in \cref{LOfig:fig2}d. In both \cref{LOfig:fig2}c and d, the experimental translational diffusivity is greater than the simulated translational diffusivity, while the experimental data shows slightly larger
fluctuations because of the experimental uncertainties. We have observed this difference before for flexible colloidal chains and it can likely be ascribed to the fact that in experiments, the substrate is a hydrogel with a finite slip length, while the simulations assume a no-slip boundary condition.\cite{Verweij2021chains}

Next, we consider the rotational diffusivity of the loops, which is defined as the in-plane rotation of the $x$-axis shown in \cref{LOfig:fig2}a or, equivalently, rotation of the cluster around the out of plane axis, for a plane parallel to the substrate.  From \cref{LOfig:fig2}e, we see that the experimentally measured rotational diffusivity is slightly lower than the simulated rotational diffusivity, but both show the same dependence on shape. Specifically, the more compact square configuration has a higher rotational diffusivity than the more extended diamond structure, as can be expected based on the greater projected surface area of the diamond structure. 

Finally, from \cref{LOfig:fig2}f, we conclude that also the flexibility of the tetramer loop depends on its shape, although more weakly: the flexibility is somewhat larger for square configurations than for diamond configurations. This indicates that more open structures have a higher flexibility, as we have also observed for chains\cite{Verweij2020, Verweij2021chains}. For chains of CSLBs, we have found that the experimental flexibility is 75 to \SI{80}{\percent} of the flexibility of the simulated chains, which is probably caused by friction of the DNA linker patch not taken into account in the simulations\cite{Verweij2021chains}. For the tetramer loops, however, we observe a drastically lower flexibility, namely, the experimental flexibility is just \SI{20\pm4}{\percent} of the simulated one, as shown in \cref{LOfig:fig2}f. This indicates that the tetramer loops experience more inter-particle friction compared to the tetramer chains. Indeed, there is one extra bond in the tetramer rings compared to the tetramer chains, which will not only lead to an increase in inter-particle friction, but also imposes an additional constraint.  We will discuss this difference in more detail in the next section.

\subsection{Shape-averaged diffusivity of rings and chains}\label{LOsec:diffusion}

\begin{figure*} \centering     \includegraphics{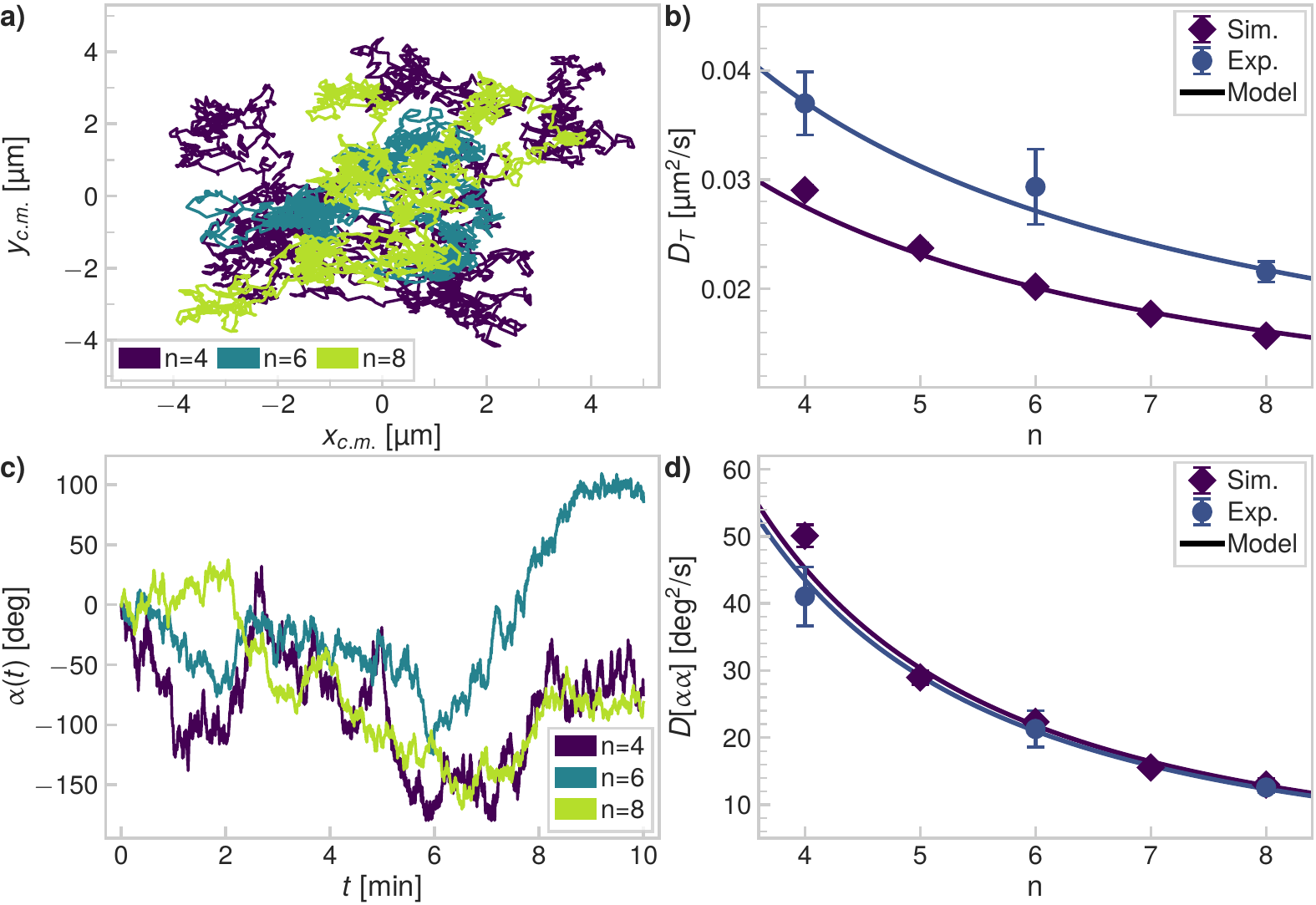}
    \caption{\textbf{Shape-averaged translational and rotational short-time diffusivity.}
\textbf{a)} The position of the center of mass of three experimental \SI{10}{\min} measurements of a tetramer, hexamer and octamer ring. \textbf{b)} Translational diffusivity of the rings as function of ring size $n$. The experimental diffusivity is greater than for the simulated data, both follow the same scaling described by \cref{eq:DT}. \textbf{c)} The orientation $\alpha$ with respect to the initial orientation of three experimental rings as function of time. \textbf{d)} The rotational diffusivity of rings as function of $n$ shows the same behavior for the experimental and simulated data, which can be described by \cref{eq:Drot}.\label{LOfig:fig5}} \end{figure*}

\subsubsection{Translational and rotational diffusivity}\label{sec:trans_rot_diff}

Having characterized the diffusivity of the tetramer loops, it is interesting to determine how the diffusivity scales with loop size $n$. A detailed shape-dependent study of the diffusivity is not feasible due to the fast increase in degrees of freedom with $n$, and hence we instead consider the shape-averaged, short-time diffusivity of the colloidal rings. In \cref{LOfig:fig5}a, we show the center of mass (c.m.) position for a tetramer, hexamer and octamer loop, as tracked from \SI{10}{\minute} experimental measurements. From this, we extract the short-time translational diffusivity $D_{T}$ as function of ring size $n$, see \cref{LOfig:fig5}b. We note that the experimental diffusivity is slightly larger than in the simulations, as we had observed previously for the tetramer loops and the flexible chains\cite{Verweij2021chains}, probably due to the different boundary conditions in experiments and simulations. We hypothesize this is due to the no-slip boundary condition on the substrate that is used in the simulations, while in the experiments the substrate is a hydrogel that has a finite slip length.

Secondly, the translational diffusivity of rings decreases as function of their size, as is expected, which we have previously shown for flexible chains\cite{Verweij2021chains}. In fact, based on Kirkwood-Riseman theory,\cite{riseman1950intrinsic} the translational diffusivity $D_{T}$ of both chains and loops are expected to scale as 
\begin{align} D_{T}
        &\propto 1 / R_{g} \propto n^{-\nu},
\label{eq:DT} \end{align} 
where the radius of gyration $R_{g}$ is defined in \cref{eq:Rgnscaling}. By fitting the simulated data in \cref{LOfig:fig5} using \cref{eq:DT}, we indeed find great agreement between model and data for $\nu=\num{0.771\pm0.003}$, the value we have found in \cref{LOsec:chainsandrings}. Deviations from the model are slightly greater in the experimental data of the colloidal rings. Nonetheless, both the experimental and simulated data are in agreement with the expected Kirkwood-Riseman scaling.

Comparison of the exponent $\nu$ we have found with measurements on the 2D diffusion of cyclic polymers is intrinsically difficult, because for the polymers, such an experiment requires adsorption of the molecules to an interface. The presence of surface asperities hinders free diffusion of cyclic polymers on the surface due to their ring-like topology and leads to a Rouse scaling law. In this case, the translational diffusion coefficient relates to the molecular weight, $M_n\propto n$, as $D_T\propto M_n^{-\mu}$,  where $\mu$ was found to be $\mu=0.75$ for short ($n<70$) cyclic polymers and $\mu=1$ for long cyclic polymers\cite{mukherji_scaling_2008}.  Experiments on high $M_w$ cyclic polystyrene molecules in a good solvent and adsorbed to silica quartz confirmed $\mu=1.00\pm 0.10$ \cite{ye_interfacial_2016}. In our colloidal model system, adhesion to the surface is not necessary because gravity provides a quasi-2D confinement. Still, we find $\nu=\num{0.771\pm0.003}$ close to the value for $\mu$.    

While the diffusivity of both rings and chains exhibits the same scaling, the ratio of their diffusion coefficients is predicted to be greater than unity, and independent of $n$. For long polymers, the ring-to-chain diffusivity ratio $K\equiv D_{T, ring} / D_{T, chain}$ is predicted to be approximately equal to $K=3\pi/8\approxeq 1.2$ based on Kirkwood-Riseman theory.\cite{fukatsu1966hydrodynamic,hegde2011conformation} Renormalization group calculations  predicted $K=e^{3/8}=1.45$.\cite{Schaub1987} Reported experimental values for synthetic polymer solutions vary between $K=1.11-1.2$\cite{duval_hydrodynamic_1985, hodgson_dilute_1991, higgins_studies_1983} and $K=1.36$;\cite{griffiths_role_1995} for short plasmid DNA $K=1.24$\cite{Voordouw1978} and for longer single DNA molecules $K=1.32$ was found.\cite{Robertson2006}
In order to obtain the ratio $K$ based on our data, we calculate the ratio of the model fits for chains and rings for $n=4$ to 8. For the simulated data, we find that on average $K=\num{1.04\pm0.01}$, whereas the experimental data yields $K=\num{1.08\pm0.01}$. These values are close to the predicted ratio based on Kirkwood-Riseman theory and similar to measurements on synthetic polymer solutions. As expected, the rings show a slightly greater diffusivity, which is explained by their smaller radius of gyration compared to chains of the same number of particles.

In the same fashion as for the translational diffusivity, we can determine the shape-averaged, short-time rotational diffusivity by calculating a rotation angle as function of time. Here, we calculate the angle $\alpha$ from the rotation of the $\hat{x}$-vector defined by \cref{LOeq:xaxis} with respect to $x$-axis of the lab frame, as depicted schematically in \cref{LOfig:fig2}a for the tetramer loop. \cref{LOfig:fig5}c shows how this rotation angle $\alpha$ changes as function of time for three experimental measurements of a tetramer, hexamer and octamer loop. Based on the mean squared angular displacements of $\alpha$, the rotational diffusivity can be calculated in turn.

The rotational diffusivity is shown in \cref{LOfig:fig5}d, from which it can be seen that the experimental values are very close to the simulated ones. Analogously to the translational diffusivity, the rotational diffusivity follows the same scaling as for chain-like objects, given by\cite{riseman1950intrinsic} 
\begin{align} D[\alpha\alpha] &\propto \frac{\ln\left(2L/b\right)}{L^3}, \label{eq:Drot} \end{align} 
where $b$ is the Kuhn length (approximately equal to the bond length) and $L=b (1+(n-1)^{\nu})$, with $n$ the number of particles and $\nu$ the Flory exponent. Indeed, in \cref{LOfig:fig5}e we find a good fit of the experimental and simulated data using \cref{eq:Drot} for $\nu=\num{0.771\pm0.003}$, the value we have found in \cref{LOsec:chainsandrings}.

Lastly, we compare the ratio of rotational diffusivities of rings and chains for the available data of chains\cite{Verweij2021chains} and rings. We use a method analogous to the determination of $K$,  the ratio of translational diffusivities,  and calculate the ring-to-chain rotational diffusivity ratio $K_r\equiv D[\alpha\alpha]_{ring} / D[\alpha\alpha]_{chain}$. For the simulated data, we find that on average $K_r=\num{2.18\pm0.10}$, whereas the experimental data yields $K_r=\num{1.86\pm0.01}$.

In summary, the smaller average radius of gyration of flexible rings compared to chains of the same number of particles has implications for both their translational and rotational diffusivity. For both diffusion constants, the rings have a higher diffusivity as expected and this effect is most pronounced for rotational diffusion, where rings exhibit a rotational diffusion that is approximately twice as high as that of chains. The ratio that we find for the translational diffusion of rings and chains is close to the theoretically expected value of approximately 1.2 based on Kirkwood-Riseman theory.\cite{fukatsu1966hydrodynamic,hegde2011conformation} Therefore, the Kirkwood-Riseman model is found to be an excellent model for predicting the shape-averaged short-time diffusivities of flexible colloidal rings and chains.

\subsubsection{Flexibility or conformational diffusion}

\begin{figure} \centering \includegraphics{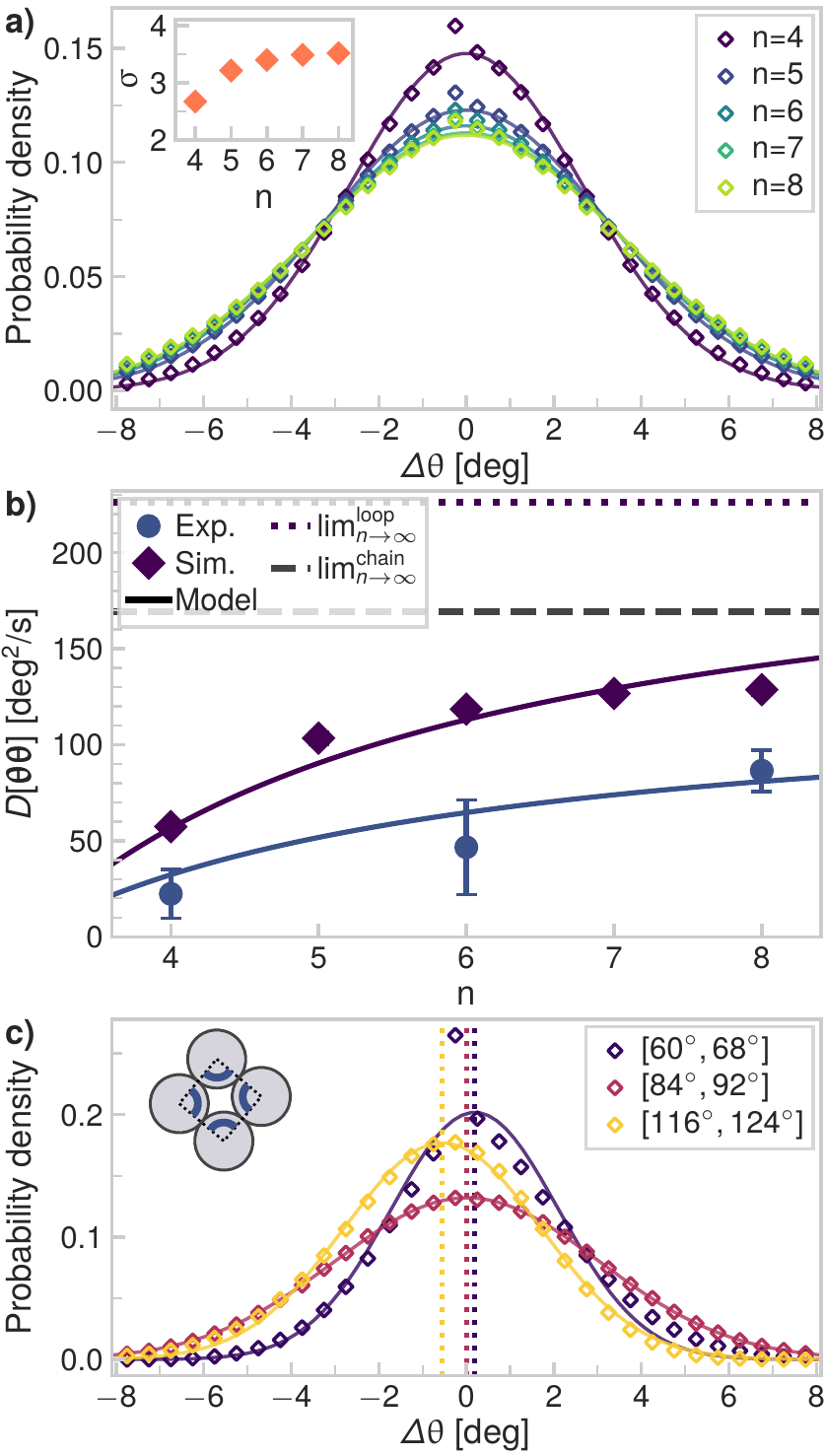}
\caption{\textbf{Shape-averaged flexibility.} \textbf{a)} Probability density of the angular displacements $\Delta\theta$ of all loops (simulated data) averaged over all opening angles $\theta$. Solid lines show a fit of a Gaussian. The inset shows the standard deviation $\sigma$ of $\Delta\theta$ vs. $n$. \textbf{b)} Flexibility of the rings as function of $n$. Solid lines are fits of \cref{eq:flextheo}. Dotted lines show the expected flexibility of chains and loops for $n \to \infty$. \textbf{c)} Probability density of the angular displacements $\Delta\theta$ of the tetramer loops (simulated data), binned per opening angle $\theta$ (see legend). Dotted lines show the mean angular displacement, solid lines show a fit of a Gaussian.\label{LOfig:fig5_flex}} \end{figure}

Besides translation and rotation, flexible colloidal objects feature another diffusion coefficient stemming from changes in their conformation.\cite{Wegener1982, Harvey1983} 
For rings of size $n$ the number of conformational degrees of freedom is given by $n-3$ based on Maxwell counting.\cite{maxwell1864calculation} It is therefore not feasible to characterize the displacements in terms of the initial shape for larger values of $n$ as we have done previously for chains with 3 and 4 segments.\cite{Verweij2020,Verweij2021chains}
Instead, we calculate the shape-averaged probability distributions of $\Delta\theta$ for all ring sizes from the simulated data (see \cref{LOfig:fig5_flex}a). As the ring size $n$ increases, the distribution becomes slightly broader. This is expected, because the confinement of the motion by the constraint of the closed ring is reduced, the larger the ring becomes. We characterize this broadening by measuring the standard deviation $\sigma$ of the distribution, whose surprising non-linear dependence with $n$ is shown in the inset. 

From the angular displacements of the flexible loops, we can also determine their conformational diffusivity $D[\bm{\theta\theta}]$, or flexibility, from the slope of the mean squared angular displacement as given by \cref{LOeq:flexibility}. From \cref{LOfig:fig5_flex}b, we can observe that for both the simulated and experimental data, the flexibility indeed increases as function of loop size $n$, as we had expected based on \cref{LOfig:fig5_flex}a. The experimental flexibility is lower in magnitude than the simulated one, which is likely caused by interparticle friction stemming from the DNA linker patch between the particles. This friction is not modeled in the simulations and therefore the simulated flexibility is higher, as we had observed previously for colloidal chains.\cite{Verweij2021chains}

The conformational diffusion coefficient $D_n[\bm{\theta\theta}]$ 
shows a non-linear dependence with increasing ring size. This can be understood from that the standard deviation and $D_n[\bm{\theta\theta}]$ are related through $D_n[\bm{\theta\theta}]=\frac{\langle \lvert\bm{\Delta\theta}\rvert^2 \rangle}{2n\tau}$, where $\langle \lvert\bm{\Delta\theta}\rvert^2 \rangle\propto \sigma_n^2$. While these data have been averaged over all internal opening angles, averaging is not the cause for the non-linear behavior, as we find the same behavior for the conformational diffusion coefficient of a single opening angle in the ring. 

Interestingly, the chain length above which $D_n[\bm{\theta\theta}]$ starts to saturate coincides with the chain length above which the steric constraints in the opening angle are lifted, and the free energy differences between small and large opening angles decreases again, see \cref{LOfig:fig2}. At the same time, the number of conformational degrees of freedom increases and the distribution of the radius of gyration broadens. Combined, this leads to a greater range of possible displacements that are sterically allowed and therefore, a broader probability distribution for the larger ring sizes.

We therefore hypothesize that the constraints stemming from the requirement that the structures need to remain a closed ring are causing this dependence on $n$. This ring constraint imposes that all displacements $\Delta\theta_i$ of the internal opening angles add up to zero between two time steps of lag time $\tau$. More importantly, the $n-3$ conformational degrees of freedom require that only $n-3$ of these displacements can be chosen independently. However, the remaining $3$ displacements then are determined by the requirement of the closed ring topology. If we consider the independent displacements to have a distribution with standard deviation $\sigma$, and the dependent displacements to have a distribution with standard deviation $\sigma_{\mathrm{fix}}$, we can write:
$\langle \lvert\bm{\Delta\theta}\rvert^2 \rangle= (n-3)\sigma^2 + 3 \sigma^2_{\mathrm{fix}}$
However, as the three dependent displacements are completely determined by the choice of the independent displacements, their distribution has an effective standard deviation $\sigma_{\mathrm{fix}}=0$. Thus 
\begin{align}
\langle \lvert\bm{\Delta\theta}\rvert^2 \rangle= (n-3)\sigma^2
\end{align}
and we find that
\begin{align}
D_n[\bm{\theta\theta}]=\frac{\langle \lvert\bm{\Delta\theta}\rvert^2 \rangle}{2n\tau}=\frac{(n-3)\sigma^2}{2n\tau}\label{eq:flextheo}
\end{align}

By fitting this prediction to our experimental data, we find $\sigma^2/(2\tau)=\SI{129\pm 16}{deg^2/s}$ and for the simulated data we find $\sigma^2/(2\tau)=\SI{226\pm 4}{deg^2/s}$ (indicated by the dotted line in \cref{LOfig:fig5_flex}b), which is the value that the flexibility saturates at for large $n$. The ratio between the experimental and simulated flexibility is therefore \num{0.57\pm 0.07}. This ratio is  close to the value of \numrange{0.75}{0.8} that we had found previously for colloidal chains.~\cite{Verweij2021chains} This suggest that the decrease in experimental flexibility is approximately the same for chains and loops. Most likely, it it causes by friction that is not accounted for in the simulations.

Our description of the flexibility of colloidal loops in \cref{eq:flextheo} predicts that for large $n$, the flexibility should approach a limiting value of $\sigma^2/(2\tau)$. Interestingly, for large ring sizes, the flexibility does not depend on $n$ anymore, just as the flexibility of colloidal chains is independent of the chain size. Therefore, it is instructive to compare the values of $\sigma^2/(2\tau)$ that we have found for colloidal rings to those we had determined previously for colloidal chains. For chains, we have found an average flexibility of \SI{168\pm4}{deg^2\per\second} (indicated by the dashed line in \cref{LOfig:fig5_flex}b) for our simulated data and \SI{100\pm20}{deg^2\per\second} for the experimental data.\cite{Verweij2021chains} Therefore, we find that the ratio between the flexibility of loops and chains is \num{1.34\pm 0.03} for the simulated data and \num{1.30\pm 0.30} for the experimental data. These values agree within error. They indicate that for large $n$, the flexibility of colloidal loops is greater than that of colloidal chains.

A possible explanation for the prediction that colloidal loops have a greater conformational flexibility than colloidal chains as $n \to \infty$ might be found in the differences of their translational diffusivity. 
For ring-structures, the  translational diffusivity is larger by a factor $K$ of \numrange{1.1}{1.4}, as discussed in \cref{sec:trans_rot_diff}. Indeed, the ratio between the flexibility of loops and chains that we find here also falls into this range of values for K. Therefore, it is likely that the larger flexibility of the colloidal rings as $n \to \infty$ is a result of their greater translational diffusivity compared to colloidal chains of the same size. We note that the translational diffusivity has also previously been identified to be important for the flexibility, where it was found that the maximum value of the flexibility was determined by the translational diffusion coefficient of the individual spheres.\cite{McMullen2018,Verweij2021chains} 

Finally, we also consider effects that stem from excluded volume interactions and hydrodynamics. To do so we consider the simplest ring that has just a single degree of freedom, the tetramer loop. In \cref{LOfig:fig5_flex}c we show the probability density function of displacements $\Delta\theta$ during a lag time of $\tau=\SI{0.05}{\second}$ for the simulated data for three initial configurations.  First, for the square configuration ($\theta\in[\SI{84}{\degree},\SI{92}{\degree}]$), we find that as expected displacements are symmetrically distributed around zero. This is in contrast to the diamond configurations: for $\theta\in[\SI{60}{\degree},\SI{68}{\degree}]$ we find a small bias towards positive displacements and for $\theta\in[\SI{116}{\degree},\SI{124}{\degree}]$ we find the opposite, a bias towards negative displacements. The origin of these shifts likely stems from a combination of the constraint that the internal opening angle cannot be smaller than \SI{60}{\degree} or larger than \SI{120}{\degree}, and hydrodynamic interactions when particles that span the internal opening angle are close to each other.\cite{Verweij2020}

In addition to the mean value, the distribution for square configurations is also broader than the distribution of the diamond configurations. This leads to the slightly larger diffusivity for the square configurations that we have observed in \cref{LOfig:fig2}f for the short-time diffusivity of tetramer loops. Therefore, we hypothesize that the differences in flexibility as function of angle $\theta$ that we have observed in \cref{LOfig:fig2}f are in part caused by steric constraints. Other contributions stem from hydrodynamic interactions between the particles. 

To summarize, we have found that the flexibility of the colloidal loops increases as function of loop size $n$. As $n \to \infty$, the flexibility saturates to a higher value than the flexibility of colloidal chains. For smaller loop sizes, steric restrictions lead to a lower flexibility. We have proposed a scaling relation in terms of the number of conformational degrees of freedom and find that it describes our simulated and experimental data well. The observed scaling could have implications for other microscopic ring like structures found in biology, such as ring polymers. Additionally, the observation that in addition to hydrodynamic properties, steric constraints can have a large effect on the conformational diffusivity of flexible objects as well could have broader implications beyond ring-like structures.

\section{Conclusions}

In conclusion, we have studied a model system of flexible colloidal loops using experiments and simulations, to gain greater insight in the physical processes that govern the behavior of reconfigurable ring-like microscopic objects, such as ring polymers. First, we have found that for the
loops consisting of four to eight spheres, that steric constraints have a large effect on the possible conformations of the rings, as characterized by their opening angles and radius of gyration. Indeed, we have found differences in free energy of around $6~k_{B}T$ between the most and least likely
configurations of the rings. We found that the scaling of the radius of gyration scales according to Flory theory, and that the ratio of radii of gyration of rings and chains is close to the expected value derived for molecular polymers by renormalization theory.

The diffusive properties of the rings show the same scaling as function of size as flexible colloidal chains and follows predictions based on Kirkwood-Riseman theory. We find that the ratio between the translational diffusivity of rings and chains is close to the expected theoretical value and observe that the rotational diffusivity of rings is approximately twice as high as that of chains. In contrast to chains, for which the flexibility is nearly constant, the flexibility, or conformational diffusivity, of rings increases as function
of ring size up to a limiting value, set by hydrodynamic friction. We propose a simple scaling that models that considers the number of conformational degrees of freedom in the displacements and found that the proposed scaling is in good agreement with the experimental and simulated data.

This observed scaling could have broader implications for other microscopic ring like structures found in synthetic and biological systems, such as ring polymers. In particular, our observation that topological constraints can have a large effect on the conformational diffusivity of flexible objects as well could have broader implications beyond ring-like structures. 
Therefore, it would be interesting to investigate the conformational diffusivity of (model systems for) structures where geometric constraints play a significant role, such as in floppy colloidal crystals and intrinsically disordered
proteins.

\subsubsection*{Acknowledgments} 

We thank Ali Azadbakht for the design and setup of the Optical Tweezers and his technical support. We are grateful to Aleksandar Donev and Brennan Sprinkle for fruitful discussions and for providing us with example code for the simulations. We thank Yair Shokef and Martin van Hecke for useful discussions. The simulations were partly performed using the ALICE compute resources provided by Leiden University.  This project has received funding from the European Research Council (ERC) under the European Union's Horizon 2020 research and innovation program (grant agreement no. 758383).

\bibliography{references}

\end{document}